\documentclass[12pt]{iopart}      
\begin{document}
\title{Chiral anomaly for V-A fields in four- and six-dimensional curved space}
\author{S. Yajima, K. Eguchi, M. Fukuda, T. Oka, H. Taira and S. Yamashita}
\address{Department of Physics, Kumamoto University, \\
2-39-1 Kurokami, Chuo-ku, Kumamoto 860-8555, Japan\\
E-mail: yajima@sci.kumamoto-u.ac.jp}
\begin{abstract}
The chiral U(1) anomalies associated 
with a fermion of spin $\frac{1}{2}$ interacting 
with nonabelian vector and axial-vector fields 
in four- and six-dimensional curved space are given in tensorial form. 
\end{abstract}
%
%

\section{Introduction}
The chiral U(1) anomaly has been derived  
by calculating some Feynman triangle diagrams of fermions 
in four-dimensional quantum electrodynamics~\cite{Adler,BJ}  
and is studied in quantum field theory 
because it is a fruitful topic. 
The anomaly is obtained from the chiral transformation of 
the Euclidean path integral measure for gauge theories
with fermions~\cite{Fujikawa1,Fujikawa2}. 
The derivation of the anomalous axial-vector
Ward--Takahashi identities in the method 
has attracted some attention~\cite{FS}.
The anomaly is related to the chiral magnetic effect and
topological insulators in condensed
matter physics~\cite{CME1,CME2,CME3,CME4}.     
 
In quantum field theory, 
several fermionic loop corrections are perturbatively described 
by a fermion propagator with a background field. 
The chiral U(1) anomaly in four-dimensional curved space~\cite{Kimura} 
has been obtained using the heat kernel~\cite{Schwinger},
by which the propagator should be given.
Some anomalies for chiral fermions interacting with
gauge fields in higher even-dimensional curved space 
can be calculated using this method.
The chiral U(1) anomaly in six dimensions 
has topologically similar features 
to that in ten dimensions 
because of the index theorem~\cite{IndexTheorem1,IndexTheorem2}.    

In $d=10$, $N=1$ supergravity~\cite{SG1,SG2}, 
the Rarita--Schwinger field $\psi_{\mu}$ 
describes a gravitino by fixing a gauge suitably, 
and the evaluation of the anomalies requires 
the heat kernel for a fermion with spin $\frac{3}{2}$~\cite{GF1,GF2,GF3,GF4,GF5}.
However, by regarding the vector index ``$\mu$'' of $\psi_{\mu}$ 
as that of the representation matrices of gauge group SO(10), 
the heat kernel for a spinor with
spin $\frac{3}{2}$ can be expressed by that for a spinor 
with spin $\frac{1}{2}$. 
Moreover, the Lagrangian in the supergravity contains not only the
minimal interaction between the gravitino and the gravitational fields 
but also four-gravitino interactions. 
In the fermionic one-loop diagram, the four-gravitino vertex 
is connected by two internal and two external lines. 
Then, by regarding the two fermion external lines as a boson line, 
the four-gravitino interaction is treated as two-gravitino interactions 
with bosonic background fields, which are odd-order tensors. 
The totally antisymmetric tensor of the highest order $2n-1$ 
in even $2n$ dimensions
is rewritten as the axial-vector 
by contracting the Levi--Civita symbol with its tensor.

A fermion on which the projection matrix $(1 \pm \gamma_{2n+1})/2$ acts
is expressed by the Weyl spinor with either positive or negative
chirality, 
which corresponds to an eigenstate with right- or left-handed
helicity in massless fermions, respectively. 
In the Dirac operator of the Weyl fermion, the gauge fields 
are separated into two types of boson by the projection. 
However, we must note the Hermiticity of the bosonic coupling 
in the Dirac operator ${\ooalign{\hfil/\hfil\crcr${D}$}}$  
because, to evaluate the correctness 
of the result of the anomaly, it is necessary to calculate 
by replacing ${\ooalign{\hfil/\hfil\crcr${D}$}}^2$ with 
${\ooalign{\hfil/\hfil\crcr${D}$}}
{\ooalign{\hfil/\hfil\crcr${D}$}}^{\dag} 
+ {\ooalign{\hfil/\hfil\crcr${D}$}}^{\dag}
{\ooalign{\hfil/\hfil\crcr${D}$}}$\,; 
if the Dirac operator is not Hermitian, 
${\ooalign{\hfil/\hfil\crcr${D}$}}^{\dag} 
\neq {\ooalign{\hfil/\hfil\crcr${D}$}}$. 
In contrast, the gauge bosons in
${\ooalign{\hfil/\hfil\crcr${D}$}}$ and  
${\ooalign{\hfil/\hfil\crcr${D}$}}^{\dag}$ 
are described by a suitable linear combination 
of the vector  (V) and axial-vector (A) fields. 
The chiral U(1) anomaly for the Weyl fermion is half of that for the
Dirac fermion, 
up to the sign of the chirality of the Weyl fermion. 
Therefore, it is simple to begin the derivation of the anomaly 
with the Hermitian gauge couplings in the Dirac operator. 
In this article, we consider 
the chiral U(1) anomalies for the Dirac fermion 
of spin $\frac{1}{2}$ interacting 
with Hermitian and nonabelian V-A fields 
in four- and six-dimensional curved space. 

\section{Heat kernel}
The heat kernel $K^{(d)}(x,x')$ for a fermion 
of spin $\frac{1}{2}$ in $d$ dimensions is defined by 
\begin{eqnarray}
 {\partial \over \partial t}\, K^{(d)}(x,x';t) =
-\, H K^{(d)}(x,x';t), \label{eq:K-eq} \\
 K^{(d)}(x,x';0) =
{\bf 1}\, |h(x)|^{- \frac{1}{2}}\, 
|h(x')|^{- \frac{1}{2}}\, \delta^{(d)}(x,x'), \label{eq:K-con} 
\end{eqnarray}
where $\delta^{(d)}(x,x')$ 
is the $d$-dimensional invariant $\delta$ function, 
${\bf 1}$ is the unit matrix for the spinor, 
and $h=\det{h^a{}_\mu}$, in which $h^a{}_{\mu}$ 
is a vielbein in curved space. 
Here $H$ is the second-order differential operator 
corresponding to the square of the Dirac operator 
${\ooalign{\hfil/\hfil\crcr${D}$}}$ for the fermion $\psi$, 
\begin{eqnarray}
\fl \qquad && H = {\ooalign{\hfil/\hfil\crcr${D}$}}^2 
= D_{\mu} D^{\mu} + X, \quad 
{\ooalign{\hfil/\hfil\crcr${D}$}}
= \gamma^{\mu} \nabla_{\mu} + Y, \quad 
D_{\mu} = \nabla_{\mu} + Q_{\mu},
\quad 
Q_{\mu} = \frac{1}{2} \{ \gamma_{\mu} , Y \},  
\nonumber \\
\fl \qquad && X = Z - \nabla_{\mu} Q^{\mu} - Q_{\mu} Q^{\mu}, \quad
\nabla_{\mu} \psi = \partial_{\mu} \psi
+ \frac{1}{4}\, \omega^{ab}{}_{\mu}\, \gamma_{ab}\, \psi , 
\quad \gamma_{a_1 \cdots a_j} = \gamma_{[a_1} \cdots \gamma_{a_j]},
\nonumber \\
\fl \qquad  && Z = \frac{1}{2}\, \gamma^{\mu\nu}\, [ \nabla_{\mu}, \nabla_{\nu} ] 
+ \gamma^{\mu} \nabla_{\mu} Y + Y^2, \qquad 
[ D_{\mu}, D_{\nu} ]\psi = \Lambda_{\mu\nu}\, \psi ,
\end{eqnarray}
where $\omega^{ab}{}_{\mu}$ is Ricci's coefficient of rotation. 
When in $d=2n$ dimensions
the fermion interacts with vector and
axial-vector fields that do not commute,  
the Dirac operator contains the coupling 
of these bosons in $Y$: 
\begin{equation}
\fl \qquad Y = \gamma^{\mu} V_{\mu} + \gamma_{2n+1} \gamma^{\mu} A_{\mu},\quad
V_{\mu}\equiv V^a_{\mu} T^a, \quad A_{\mu}\equiv A^a_{\mu}T^a, 
\quad \gamma_{2n+1} = i^n \gamma^1 \gamma^2 \cdots \gamma^{2n}. 
\end{equation}
Here $T^a$ is the representation matrix of a gauge group, and 
$V^a_{\mu}\ (A^a_{\mu})$ is purely imaginary (real), 
because of the Hermiticity of the Dirac operator. 
The quantities $Q_{\mu}$, $X$, and $\Lambda_{\mu\nu}$ 
are expressed in the following tensorial form: 
\begin{eqnarray}
\fl \qquad &&Q_{\mu} = V_{\mu} - \gamma_{2n+1}\, \gamma_{\mu\rho} A^{\rho}, \qquad
F_{\mu\nu} = \partial_{\mu} V_{\nu} - \partial_{\nu} V_{\mu}
+ [V_{\mu} , V_{\nu} ], \nonumber \\
\fl \qquad &&X = -\, {1 \over 4}\, R + 2(n-1) \, A_{\mu} A^{\mu} 
- \gamma_{2n+1}\, A^{\mu}{}_{:\mu}
+ \gamma^{\mu\nu} \left( {1 \over 2}\, F_{\mu\nu} 
+ {2n-3 \over 2}\, [A_{\mu} , A_{\nu} ] \right), \nonumber \\
\fl \qquad &&\Lambda_{\mu\nu} 
= {1 \over 4}\, \gamma^{\rho\sigma} R_{\rho\sigma\mu\nu}
+ F_{\mu\nu} - [A_{\mu} , A_{\nu} ] 
- 2\, \gamma_{\mu\nu} A_{\rho} A^{\rho} 
+ 2\, \gamma_{[\mu|}{}^{\rho} \{ A_{|\nu]} , A_{\rho} \} \nonumber \\
\fl \qquad &&\qquad\quad +\, 2\, \gamma_{2n+1}\, \gamma_{[\mu|\rho}  
A^{\rho}{}_{:|\nu]}
- 2 \, \gamma_{\mu\nu\rho\sigma} A^{\rho} A^{\sigma}, 
\label{Lam}
\end{eqnarray}
where $R_{\alpha\beta\mu\nu}$ stands for the curvature tensor,
and the colon ($:$) represents
the Riemannian covariant differentiation $\nabla_{\mu}\,$.
The completely antisymmetric product
$\gamma_{\mu\nu\rho\sigma}$ of $\gamma$ matrices 
in the last term of $\Lambda_{\mu\nu}$ 
is rewritten as
$- \epsilon_{\mu\nu\rho\sigma} \gamma_5$ 
and $- \frac{i}{2} \epsilon_{\mu\nu\rho\sigma\kappa\lambda} 
\gamma_7 \gamma^{\kappa\lambda}$ 
in four and six dimensions, respectively.

The differential equation $(\ref{eq:K-eq})$ of the heat kernel 
for the fermion interacting 
with the general boson fields is not strictly solvable . 
Therefore, the heat kernel is usually calculated 
by using De Witt's ansatz \cite{DeWitt}, 
which automatically satisfies $(\ref{eq:K-con})$, 
\begin{equation}
 K^{(2n)}(x,x';t) \sim {\Delta^{1/2}(x,x') \over (4 \pi t)^n}\, 
{\rm exp} \left({\sigma(x,x') \over 2t} \right) 
\sum_{q=0}^{\infty} a_q(x,x')\, t^q,  
\label{eq:DWansatz}
\end{equation}
where $\sigma(x,x')$ and $\Delta(x,x')$ are 
half of the square of the geodesic distance 
and the Van Vleck--Morette determinant between $x$ and $x'$, respectively, 
and $a_q(x,x')$ are bispinors. 
Note that the coincidence limit of $a_0$ is 
$\lim_{x'\to x}a_0(x,x') \equiv [a_0](x)={\bf 1}$, 
and that the metric tensor in curved space is 
$g_{\mu\nu} = h^a{}_{\mu} h^b{}_{\nu} \eta_{ab}$ 
with $\eta_{ab} = -\, \delta_{ab}$ in flat tangent space. 

\section{Chiral U(1) anomaly}
The formal expression of the chiral U(1) anomaly 
${\mathcal A}^{(2n)}$ in $2n$ dimensions 
is derived from the path integral measure \cite{Fujikawa1,Fujikawa2} 
using the coincidence limit of the bispinor $a_n$ of the heat kernel, 
\begin{equation}
\nabla_{\mu} \langle \bar{\psi}(x) \gamma^{\mu} \gamma_{2n+1} \psi(x)
 \rangle = {\mathcal A}^{(2n)}(x)
= {2i \over (4\pi)^n }\, {\rm Tr} 
\left( \gamma_{2n+1}\, [a_n] (x) \right),  
\end{equation}
where ${\rm Tr}$ runs over both indices of the $\gamma$ matrices 
and representation matrices $T^a$ of the gauge group.
The concrete form of $[a_n]$ is given as follows \cite{Gilkey}:  
\begin{eqnarray}
\fl \qquad && [a_2] = {1 \over 12} {\Lambda}_{\mu\nu} {\Lambda}^{\mu\nu}
+ {1 \over 6} X_{!\mu}{}^{\mu}
+ {1 \over 2} \left({1 \over 6} R + X \right)^2 
+ \cdots , \label{a2} \\
\fl \qquad && [a_3] = {1 \over 60} \Big( 
-{1 \over 3} \left(X_{!\mu}{}^{\mu}{}_{\nu}{}^{\nu}
+ X_{!\mu\nu}{}^{\nu\mu} + X_{!\mu\nu}{}^{\mu\nu}\right)  
- {1 \over 3} {\Lambda}_{\mu\nu}{}^{!\nu}
{\Lambda}^{\mu\rho}{}_{!\rho}
- {4 \over 3} {\Lambda}_{\mu\nu!\rho}
{\Lambda}^{\mu\nu!\rho}
\nonumber \\
\fl \qquad && \phantom{[a_3] ={1 \over 60} \Big( }
- 4 {\Lambda}^{\mu\nu}
{\Lambda}_{\mu\rho}{}^{!\rho}{}_{\nu} 
- {10 \over 3} R^{\mu\nu} 
{\Lambda}_{\mu\rho} {\Lambda}_{\nu}{}^{\rho}
+ R^{\mu\nu\rho\sigma} {\Lambda}_{\mu\nu} {\Lambda}_{\rho\sigma}
- 6{\Lambda}_{\mu}{}^{\nu}
{\Lambda}_{\nu}{}^{\rho} {\Lambda}_{\rho}{}^{\mu} \Big)
 \nonumber \\
\fl \qquad && \phantom{[a_3] =}+ {1 \over 12} \bigg\{ {1 \over 6} R + X , 
- {1 \over 2} {\Lambda}_{\mu\nu} {\Lambda}^{\mu\nu}
- X_{!\mu}{}^{\mu} - {1 \over 5} R_{:\mu}{}^{\mu} \nonumber \\
\fl \qquad && \phantom{[a_3] = + {1 \over 12} \bigg\{ }
+ {1 \over 30} R_{\mu\nu} R^{\mu\nu} 
- {1 \over 30} R_{\mu\nu\rho\sigma} R^{\mu\nu\rho\sigma} \bigg\}
- {1 \over 36} [ X_{!\mu} , \Lambda^{\mu\nu}{}_{!\nu} ]
\nonumber \\
\fl \quad && \phantom{[a_3] =}- {1 \over 12} \left({1 \over 6} R + X \right)_{!\rho}
\left({1 \over 6} R + X \right)^{!\rho}
- {1 \over 6} \left({1 \over 6} R + X \right)^3 + \cdots,
\label{a3}
\end{eqnarray}
where the exclamation mark ($!$) represents
the modified covariant differentiation $D_{\mu}$, 
and some terms without a $\gamma$ matrix are omitted from (\ref{a2})
and (\ref{a3}). 
After a straightforward calculation, 
the tensorial form of the anomalies 
in four and six dimensions is obtained as
\begin{eqnarray}
\fl \qquad {\mathcal A}^{(4)} &=& 
{i \over 8 \pi^2}\, {\rm tr}\bigg[ 
\epsilon_{\alpha\beta\gamma\delta}
\left(-\, {1 \over 48} R^{\alpha\beta}{}_{\rho\sigma} 
R^{\gamma\delta\rho\sigma}
- {1 \over 2} F^{\alpha\beta} F^{\gamma\delta}
- {2 \over 3} A^{\alpha:\beta} A^{\gamma:\delta} 
\right. \nonumber \\
\fl \qquad &&\quad \left.
-\, {2 \over 3} [A^{\alpha} , A^{\beta}] F^{\gamma\delta} \right)
+ \left( -\, {2 \over 3} A_{\nu:\mu}{}^{\mu} 
+ {1 \over 3} R A_{\nu}
+ {8 \over 3} A_{\nu} A_{\mu} A^{\mu} \right)^{:\nu} \bigg]
, \label{A4} 
\end{eqnarray}
\begin{eqnarray}
\fl \qquad {\mathcal A^{(6)}} &=& 
{i \over 4\pi^3}\, {\rm tr}\bigg[ - {i \over 8}\,
\epsilon_{\alpha\beta\gamma\delta\kappa\lambda}
\left({1 \over 48} R^{\alpha\beta}{}_{\rho\sigma} R^{\gamma\delta\rho\sigma}
+ {1 \over 6} F^{\alpha\beta} F^{\gamma\delta} \right) F^{\kappa\lambda} 
\nonumber \\
\fl \qquad &&+\, \left({1 \over 180} \left(A^{\mu}{}_{:\mu\nu}{}^{\nu\rho} 
+ A^{\mu}{}_{:\mu\nu}{}^{\rho\nu} 
+ A^{\mu}{}_{:\mu}{}^{\rho\nu}{}_{\nu} \right) 
\right. \nonumber \\
\fl \qquad &&\quad +\, {1 \over 6} \left(-\, F^{\mu\nu} F_{\mu\nu} A^{\rho}
+ \{F^{\mu\nu} , F^{\rho}{}_{\nu} \} A_{\mu} \right) 
- {1 \over 72} R A_{\mu}{}^{:\mu\rho} \nonumber \\
\fl \qquad &&\quad +\, {1 \over 120} \left(-\, R^{:\rho} A_{\mu}{}^{:\mu}
+ R_{:\mu} A^{\rho:\mu} + R_{:\mu} A^{\mu:\rho} \right)
- {1 \over 90} R^{\rho\mu} A_{\mu:\nu}{}^{\nu} \nonumber \\
\fl \qquad &&\quad +\, {2 \over 45} R_{\mu\nu} A^{\mu:\nu\rho} 
+ {1 \over 30} \left(-\, R^{\rho\mu:\nu} + R^{\mu\nu:\rho} \right)
A_{\nu:\mu}
- {1 \over 36} R^{\mu\nu\lambda\rho} A_{\mu:\lambda\nu} \nonumber \\
\fl \qquad &&\quad +\, {1 \over 36} R_{\mu\nu} R^{\mu\lambda\nu\rho} A_{\lambda} 
- {1 \over 180} R^{\mu\nu\kappa\rho} R_{\mu\nu\kappa\lambda} 
A^{\lambda}
- {7 \over 1440} R^{\mu\nu\kappa\lambda} R_{\mu\nu\kappa\lambda} 
A^{\rho} \nonumber \\
\fl \qquad &&\quad \left. +\, {1 \over 90} R_{\mu\nu} R^{\mu\rho} A^{\nu} 
- {1 \over 72} R R^{\mu\rho} A_{\mu}
+ {1 \over 288} R^2 A^{\rho} \right)_{:\rho} \nonumber \\
\fl \qquad &&-\, {i \over 12}\, \epsilon_{\alpha\beta\gamma\delta\kappa\lambda}
\left(F^{\alpha\beta} A^{\gamma:\delta} A^{\kappa} \right)^{:\lambda} 
\nonumber \\
\fl \qquad &&+\, \left( {11 \over 15} A_{\mu} A^{\mu} A^{\rho:\nu}{}_{\nu}
+ {4 \over 3} A_{\mu} A^{\mu} A_{\nu}{}^{:\nu\rho} 
- {19 \over 15} A_{\mu} A^{\mu} A_{\nu}{}^{:\rho\nu} \right. \nonumber \\
\fl \qquad &&\quad -\, {1 \over 10} \{A^{\rho}, A^{\mu}\} A^{\nu}{}_{:\mu\nu} 
+ {3 \over 10} \{A^{\rho} , A^{\mu} \} A_{\mu:\nu}{}^{\nu}
+ {1 \over 30} \{A^{\rho} , A^{\mu} \} A^{\nu}{}_{:\nu\mu} \nonumber \\
\fl \qquad &&\quad +\, {1 \over 3} \{A^{\mu}, A^{\nu}\} A_{\mu:\nu}{}^{\rho} 
+ {1 \over 15} \{A^{\mu}, A^{\nu} \} A^{\rho}{}_{:\mu\nu}
- {11 \over 30}  \{A^{\mu}, A^{\nu}\} A_{\mu}{}^{:\rho}{}_{\nu}  \nonumber \\
\fl \qquad &&\quad 
- {17 \over 30} \{A_{\mu:\nu}, A^{\nu:\rho} \} A^{\mu} 
- {1 \over 30} \{A^{\rho:\mu}, A_{\mu:\nu} \} A^{\nu} \nonumber \\
\fl \qquad &&\quad +\, {4 \over 5} \{A^{\rho:\mu}, A_{\nu:\mu} \} A^{\nu} 
- {1 \over 10} \{A^{\mu}{}_{:\mu}, A^{\nu:\rho} \} A_{\nu}
+ {1 \over 15} \{A^{\mu}{}_{:\mu}, A^{\rho:\nu} \} A_{\nu}
\nonumber \\
\fl \qquad &&\quad +\, {1 \over 15} A^{\mu}{}_{:\mu} A^{\nu}{}_{:\nu} A^{\rho}
+ {2 \over 5} A^{\mu:\nu} A_{\mu:\nu} A^{\rho} 
- {2 \over 15} A^{\mu:\nu} A_{\nu:\mu} A^{\rho} \nonumber \\
\fl \qquad &&\quad -\, {3 \over 10} F^{\nu\rho} [A_{\mu} A^{\mu} , A_{\nu} ]
- {2 \over 5} F_{\mu\nu} \{A^{\mu} A^{\nu} , A^{\rho} \}
- {2 \over 15} F_{\mu\nu} A^{\mu} A^{\rho} A^{\nu} \nonumber \\
\fl \qquad &&\quad \left. 
+\, {29 \over 45} R^{\rho\mu} A_{\mu} A_{\nu} A^{\nu}
- {1 \over 45} R^{\mu\nu} A_{\mu} A_{\nu} A^{\rho}
- {1 \over 18} R A_{\mu} A^{\mu} A^{\rho} \right)_{:\rho} 
\nonumber \\
\fl \qquad &&-\, {i \over 20}\, \epsilon_{\alpha\beta\gamma\delta\kappa\lambda}
\left(A^{\alpha} A^{\beta} A^{\gamma} A^{\delta:\kappa}
\right)^{:\lambda} \nonumber \\
\fl \qquad &&+\, \left( {1 \over 5} A_{\mu} A^{\mu} A_{\nu} A^{\nu} A^{\rho}
- {1 \over 3} A_{\mu} A_{\nu} A^{\mu} A^{\nu} A^{\rho}
+ {2 \over 5} A_{\mu} A_{\nu} A^{\nu} A^{\mu} A^{\rho} \right)_{:\rho} 
\bigg], \label{A6}
\end{eqnarray}
where 
${\rm tr}$ runs only over the internal group indices.
We note that the terms of three degrees of $A_{\mu}$ in (\ref{A6}) 
can be rewritten using the identities of the curvature tensor and 
the properties of matrices, as follows:
\begin{eqnarray}
&& {\rm tr}
\left[ ( A_{\lambda} A^{\lambda} A^{\rho} )^{:\nu}{}_{\nu}
- ( A_{\lambda} A^{\lambda} A^{\nu} )^{:\rho}{}_{\nu} 
\right]_{:\rho} = 0, \nonumber \\
&& R^{\lambda\alpha\beta\rho}\,
{\rm tr} \left[ \{ A_{\alpha}, A_{\beta} \} A_{\lambda}
\right]_{:\rho} = 0, \nonumber \\
&& 
{\rm tr} \left[ A_{\lambda} A^{\lambda}
( A_{\nu}{}^{:\rho\nu} - A_{\nu}{}^{:\nu\rho} )
- R^{\nu\rho}\, A_{\nu} A_{\lambda} A^{\lambda}
\right]_{:\rho} = 0, 
\end{eqnarray}
although these terms may naively appear with an ambiguous constant factor.
Using similar identities, the anomaly terms 
may give equivalent expressions,
though those seem to be different forms.

\section{Discussion}
In this article, the first calculation of the chiral U(1) anomaly in the
nonabelian V-A model in four- and six-dimensional curved space 
was performed. 
The covariant gauge anomaly $G$ for the left-handed Weyl fermion 
is also derived by a similar method:
\begin{equation}
\fl \qquad D_{\mu} \langle \bar{\psi}(x) \gamma^{\mu} 
{1 - \gamma_{2n+1} \over 2} T^a \psi(x) \rangle 
= G^{(2n)} = -{i \over (4\pi)^n }\, {\rm Tr} 
\left(T^a \gamma_{2n+1}\, [a_n] (x) \right).  
\end{equation}
The general form of the gauge anomaly 
in the model in four-dimensional flat space is 
well known~\cite{GA1,GA2,GA3,GA4,GA5a,GA5b,GA6a,GA6b,GA7,GA8}, 
and the chiral U(1) anomaly is easily given from the gauge anomaly. 
However, in curved space, 
a new term appears containing the Riemann curvature tensor
$R^{ab}{}_{\mu\nu}$ with the axial-vector $A_{\mu}$.

The leading terms of the chiral U(1) anomaly 
in any even-dimensional curved space 
are known to be expressed by a combination 
of the Dirac genus by the curvature tensor $R^{ab}{}_{\mu\nu}$ 
and the Chern character 
by the field strength $F_{\mu\nu}$ of the vector gauge field~\cite{DC}.
The anomaly ${\mathcal A}$ can be rewritten as the total derivative 
$\nabla_{\mu} C^{\mu}$ because of the index theorem 
when $\{ {\ooalign{\hfil/\hfil\crcr${D}$}}, \gamma_{2n+1}
\}=0$~\cite{Fujikawa1,Fujikawa2,FS}. 
As is well known, 
the anomaly provides explicit chirality-breaking terms
for the gauge-invariant and general covariant current. 
Although the modified current
$J^{\mu} \equiv \bar{\psi} \gamma^{\mu} \gamma_{2n+1} \psi
- C^{\mu}$ satisfies $\nabla_{\mu} J^{\mu} = 0$, 
the current $J^{\mu}$ does not preserve these symmetries. 
If the phase factor of the anomalous Jacobian 
from the functional measure of the path integral 
leads the $\theta$ vacuum in the presence of instantons, 
the zero-mode sector of the measure behaves abnormally.
However, the anomaly terms containing the axial-vector field 
do not affect the relationship 
between the instanton and the $\theta$ vacuum
because the terms are the divergence of the covariant quantities, 
and the Pontryagin index of the terms vanishes.
In supergravity in higher even dimensions,       
the gravitino and gaugino may interact with odd-order tensors 
constituted by the fermions. 
The chiral U(1) anomaly in the model 
has similar topological properties.
One may retain some of the leading terms of the anomaly in a model 
if one wants to explain physical phenomena. 
Note that then breakdown of the chiral 
and/or other symmetries appears. 

In (\ref{A4}) and (\ref{A6}), the matrix $T^a$ is not restricted. 
When $T^a$ is traceless, all the terms containing $F_{\mu\nu}$ and
$A_{\mu}$, which are reduced to a linear $T^a$, disappear. 
If $T^a$ is abelian, the trace operation yields
the dimension number factor of the matrix;
a commutator vanishes, and 
an anticommutator of the two fields becomes twice their product.
In the special case that only $A_{\mu}$ is abelian in (\ref{A4}),
${\mathcal A}^{(4)}$ corresponds to the chiral U(1) anomaly in 
curved space with torsion, which is the three-order antisymmetric tensor
and is rewritten by the dual axial-vector 
in four dimensions~\cite{Torsion}. 
Then, the term containing
$\epsilon_{\alpha\beta\gamma\delta}\, {\rm tr} 
([A^{\alpha} , A^{\beta}] F^{\gamma\delta})$ 
in (\ref{A4}) would disappear.
(Note that the dual torsion tensor in dimensions higher than six 
is an antisymmetric tensor of order larger than three.)
When only $T^a$ in (13) is a unit
matrix, the gauge anomaly agrees with our resultant form (10), 
up to twice the dimension number of $T^a$. 
Then, there is no term containing $A_{\mu}$ of the same degree as
the spatial dimensions 
because of the property of matrices in the trace formula 
and the contraction of the Levi--Civita symbol 
with the product of $A_{\mu}$.  
In contrast, the gauge anomaly should add terms containing
the nonzero commutator of $T^a$ and $A_{\mu}$ 
to the U(1) anomaly. 

The anomaly ${\mathcal A}_{\nu}$ expresses
quantum breaking by the general coordinate transformation, 
\begin{equation}
D^{\mu} \langle T_{\mu\nu}(x) \rangle 
= {\mathcal A}_{\nu}^{(2n)} = {1 \over 2}\, {\rm Tr} 
\left[\gamma_{2n+1}\, (2\, [a_{n!\nu}] - [a_n]_{!\nu})(x) \right].  
\end{equation}
The tensorial form of the anomaly is also a total derivative
because of the structure of the gravitational anomalies. 
The last part of it contains the covariant derivative of the U(1) anomaly.
However, the explicit form of the anomaly in the model of this article
has not been shown yet
because the term obtained from $[a_{n!\nu}]$ should be 
of a nontrivial derivative form. 
A clarification of the tensorial form of the anomaly 
would be of interest 
in order to estimate the validity of the models.

\end{document}